\documentclass[10pt,draft]{dis03}
\usepackage{epsf,amsmath}

\begin{document}
\title{DETERMINATION OF THE LONGITUDINAL PROTON STRUCTURE FUNCTION {\boldmath $F_L$} AT LOW {\boldmath $Q^2$} AT HERA 
}

\author{Ewelina Maria Lobodzinska \\
DESY, Zeuthen, Germany \\
 Institute of Nuclear Physics, Cracow, Poland\\
E-mail: ewelina@mail.desy.de }

\maketitle

\begin{abstract}
\noindent An extraction of the longitudinal proton structure function $F_L(x,Q^2
)$ from H1 data at low $Q^2$ $\sim$ 1 GeV$^2$ and low Bjorken $x$ $\sim 5\cdot 10^{-5}$ is reported. 
Two methods of extracting $F_L(x,Q^2)$ are discussed.
Theoretical predictions are compared with the data.
\end{abstract}

\section{Introduction}
The longitudinal structure function $F_L$ is directly sensitive 
to the gluon content of the proton and is therefore crucial to the understanding of the proton structure and to the determination of the gluon distribution, in particular at low momentum transfer $Q^2$ and low Bjorken $x$.
For high inelasticity $y$ the $F_L$ contribution to the reduced deep inelastic scattering (DIS) cross section becomes significant. Therefore, at high $y$ the standard procedure of extracting $F_2(x,Q^2)$ from the DIS cross section, by subtracting the theoretically computed  $F_L$ contribution,  can be reversed such that the $F_L$ contribution is extracted from the measured cross section by subtracting a calculation of $F_2(x,Q^2)$.  
 

\section{Data and extraction methods}
\begin{figure}[!thb]
\vspace*{7.0cm}
\begin{center}
\includegraphics{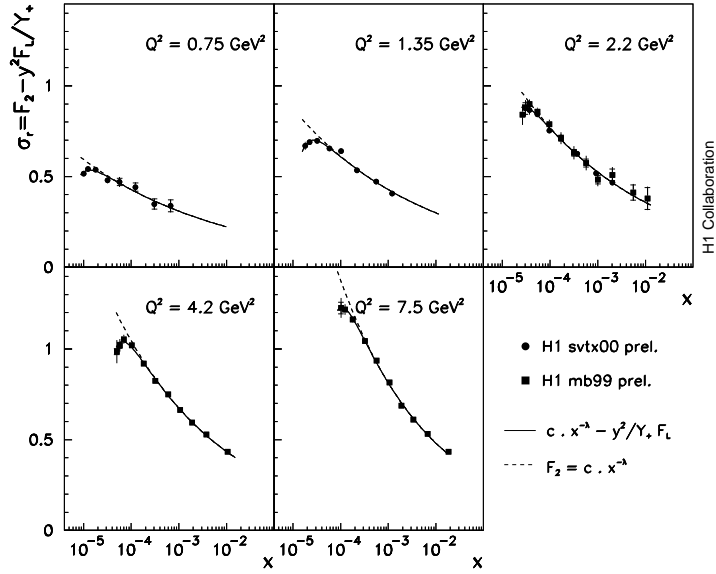}
\caption[*]{ \label{shape} The reduced cross section as a function of $x$ for different $Q^2$ bins.  The dashed lines show a function of the form $\sigma_r = c\cdot x^{-\lambda}$ representing the $F_2$ contribution to the fitted cross section. The solid lines show fits of the form $\sigma_r = c\cdot x^{-\lambda} - F_L\cdot y^2/Y_+$ , from which $F_L$ is extracted in the shape method.}
\end{center}
\end{figure}

The low $Q^2$ data used in this study were collected with the H1 detector at {\sc Hera} in a dedicated  running period in 1999 and during shifted vertex runs in 2000.
Details concerning data selection and cross section analysis can be found elsewhere~\cite{xsec}. 
The cross section, $\sigma_{r}= F_2(x,Q^2) - {y^2} \cdot F_L(x,Q^2)/(1+(1-y)^2)$, measured in various $Q^2$ bins can be seen in Fig.~\ref{shape}. 
For fixed $Q^2$, the cross section rises with decreasing $x$. However, at very low $x$ (high $y$) a characteristic bending of the cross section is observed. This occurs at all $Q^2$ values at fixed $y\sim 0.5$ and is attributed to the contribution due to the longitudinal structure function $F_L(x,Q^2)$.

\subsection{ {\boldmath $F_L$} extraction with a derivative method}
\begin{figure}[!thb]
\vspace*{6.cm}
\begin{center}
\includegraphics{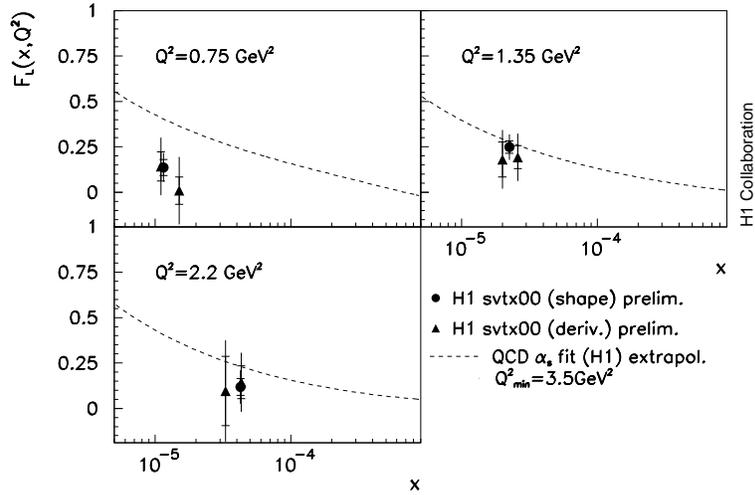}
\caption[*]{\label{shape00} Comparison of $F_L(x,Q^2)$ results, for fixed $Q^2$, from the 2000 shifted vertex data as extracted by the derivative (triangles) and the shape (points) methods, see text.
}
\end{center}
\end{figure}

The derivative method, introduced~\cite{xsec,deriv}, is based on the partial derivative of the reduced cross section calculated at fixed $Q^2$,
\begin{equation}\left(\frac{\partial\sigma_r}{\partial\ln y}\right)_{Q^2} = \left(\frac{\partial F_2}{\partial\ln y}\right)_{Q^2} - F_L\cdot y^2\cdot\frac{2-y}{{Y_+}^{2}} - \frac{\partial F_L}{\partial \ln y}\cdot \frac{y^2}{Y_+}, \label{eq_der} \end{equation} 
where $ Y_{+}=1+(1-y)^2 $.
This method assumes a linear behaviour of $\partial F_2/\partial \ln y$ with $\ln y$ and extra\-polates the information about $F_2(x,Q^2)$ from the low $y$ to the high $y$ region. It does not make full use of the information provided by the cross section measurement in the intermediate $y$ region, i.e. for the linear fit the lowest $y$ points are used but the extraction is made only for the points with highest $y$.
The result on $F_L$ consists in a few points close in $y$ with sizeable errors. The precision of the measurement does not allow to resolve the $x$ dependence of $F_L(x,Q^2)$ on this basis, see Fig.~\ref{shape00}, where the results of derivative method are shown as triangles.
Thus, a new, more precise method for the $F_L(x,Q^2)$ extraction was developed.

\subsection{{\boldmath $F_L$} extraction with a shape method}
The new, ``shape'' method employs the y-shape of the reduced cross section distribution in a given $Q^2$ bin. 
This method assumes that the shape of the reduced cross section at high $y$ (Fig.~\ref{shape}) is driven by the kinematic factor $y^2/Y_+$, and to a lesser extent by $F_L(x,Q^2)$ which, due to the narrow $x$ range covered by the  measurement at high $y$, is considered to be constant ($F_L=F_L(Q^2)$) in each $Q^2$ bin. It assumes furthermore, in agreement with previous measurements~\cite{lambda}, that the  structure function $F_2(x,Q^2)$ behaves like $x^{-\lambda}$ at fixed $Q^2$. 
On this basis the reduced cross section distribution in each $Q^2$ bin can be parametrised and fitted as  
\begin{equation}\sigma_{FIT}=c\cdot x^{-\lambda}-\frac{y^2}{1+(1-y)^2}F_L. \end{equation}
Fig.~\ref{shape} illustrates that this fit provides an excellent description of the reduced cross section in the full kinematic range. The $\lambda$ and $c$ values extracted from this fit turn out to be in good agreement with previous measurements~\cite{lambda}.

 For different $Q^2$ bins the $F_L(x,Q^2)$ points are extracted from the fit and a  bin-centre procedure is applied to obtain the correct $x$ value. 
The $F_L(x,Q^2)$ points, as obtained with the shape method, are compared with the values obtained with the derivative method in Fig.~\ref{shape00}. 
The results from both approaches are consistent. However, the errors from the shape method turn out to be significantly smaller.
Therefore, as the final result of this analysis, only the $F_L$ points extracted with the shape method are used for comparisions with theoretical calcucations.
\begin{figure}[!bht]
\vspace*{6.8cm}
\begin{center}
\includegraphics{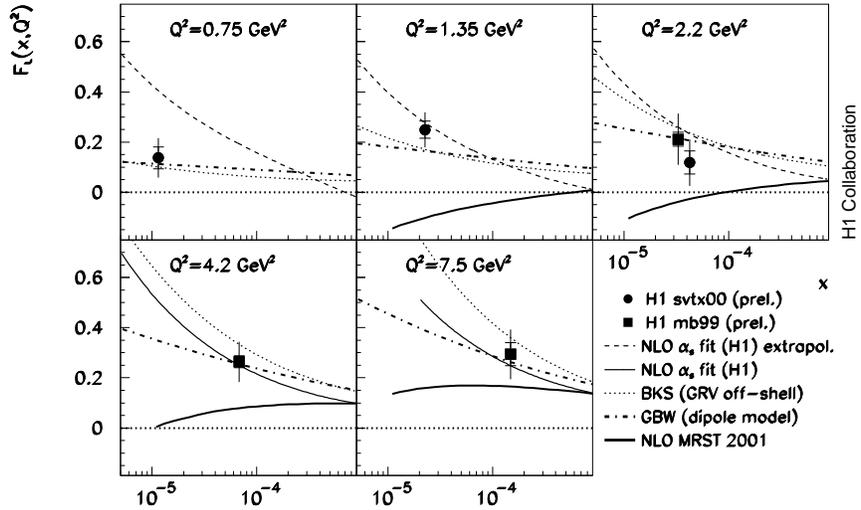}
\caption[*]{\label{fin1} $F_L(x,Q^2)$, for fixed $Q^2$, 
as extracted by the shape method. The solid, thin line represents the QCD fit to previous H1 cross section data and the dashed line the QCD fit extrapolated backwards to $Q^2$ below $Q^2_{min}$=3.5 GeV$^2$. 
Other curves show predictions of different theoretical models.
}
\end{center}
\end{figure}

\section{Results}

The longitudinal structure function $F_L(x,Q^2)$, as determined from the 1999 minimum bias and the 2000 shifted vertex H1 data, is shown in Fig.~\ref{fin1} together with predictions of various theoretical models. For all $Q^2$ bins the extracted values of $F_L(x,Q^2)$ are significantly bigger than zero.
The GBW dipole model~\cite{gbw} and the BKS (GRV off-shell) model~\cite{bks} give a good description of the extracted $F_L$ points over the whole kinematic region covered by this measurement.
The MRST 2001~\cite{mrst} NLO QCD calculation, 
however,  
undershoots the extracted points significantly. The H1 QCD fit agrees with data for higher $Q^2$, but its backward extrapolation exceeds the data for $Q^2 <$ 1 GeV$^2$.   
\begin{figure}[!bht]
\vspace*{5.8cm}
\begin{center}
\includegraphics{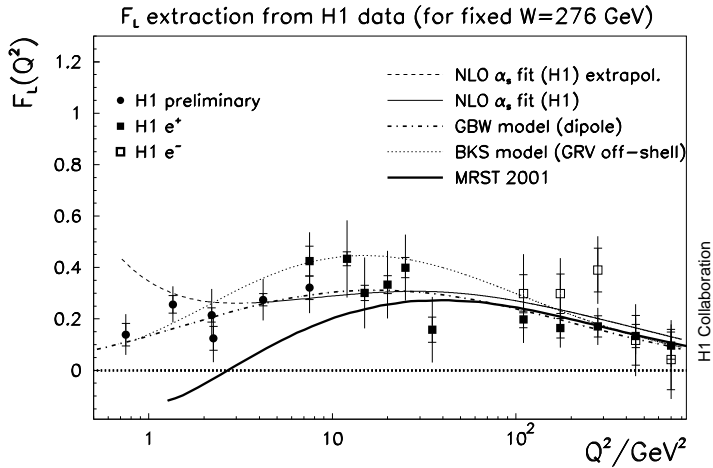}
\caption[*]{ \label{fin_all} $Q^2$ dependence of $F_L(x,Q^2)$ (at fixed W=276 GeV), summarizing the data from the H1 experiment.
}
\end{center}
\end{figure}

An overview of the current H1 data on $F_L(x,Q^2)$, from $Q^2 = 0.75$ GeV$^2$ to $Q^2 = 700$ GeV$^2$ and for fixed W=276 GeV, is given in Fig.~\ref{fin_all}. It comprises the preliminary results of the low $Q^2$ analysis described in this paper, previous results based on data collected in 96/97~\cite{deriv} and also the recently published high $Q^2$ results from $e^+p$ and $e^-p$ data~\cite{highQ2}.
The experimental points are in good agreement with the GBW dipole model~\cite{gbw} in the whole $Q^2$ range. The BKS model~\cite{bks}, which evolves steeper at low and moderate $Q^2$, is still able to describe the data.  The MRST 2001 NLO calculation~\cite{mrst} which shows a significant disagreement with the data for low $Q^2$ provides a good description only in the higher $Q^2$ region. 

\section{Summary}

Two methods of extraction of the longitudinal structure function $F_L(x,Q^2)$ from H1 inclusive cross section data in the low $Q^2$ region are presented. 
It is shown that both methods give consistent results, the shape method, however, proofs to be more precise than the derivative one. 
The measured data points are in agreement with previous results, but are more accurate  and extend the region in which $F_L$ is extracted into the very low $Q^2$ region. $F_L$ is measured to be positive even at very low $Q^2$ and $x$.
Collecting all H1 results,  $F_L(x,Q^2)$ data are presented in a wide $Q^2$ range, from 0.75 to 700 GeV$^2$. 



%
%

\end{document}